\documentclass{aa}  

\usepackage{graphicx}
\usepackage{placeins}
\usepackage{ulem}
\usepackage{txfonts}
\usepackage{natbib}

\usepackage{hyperref}
\hypersetup{
    colorlinks = true,
    linkcolor={red},
    citecolor={blue}
}
\begin{document}

   \title{Spectroscopic study of Ceres' collisional family candidates}

   \author{F. Tinaut-Ruano
          \inst{1}\fnmsep\inst{2}
          \and
          J. de Leon\inst{1}\fnmsep\inst{2}\fnmsep
          \and
          E. Tatsumi\inst{1}\fnmsep\inst{3}\fnmsep
          \and
          B. Rousseau\inst{4}
          \and
          J. L. Rizos\inst{1}
          \and
          S. Marchi\inst{5}
          }

   \institute{Instituto de Astrofísica de Canarias (IAC) \\
              Calle Vía Láctea, s/n, 38205 San Cristóbal de La Laguna, Santa Cruz de Tenerife
         \and
             University of La Laguna, Department of Astrophysics\\ Av. Astrofísico 
Francisco Sánchez, S/N, 38206 San Cristóbal de La Laguna, Santa Cruz de Tenerife
        \and
            Department of Earth and Planetary Science, University of Tokyo, 7-3-1 Hongo, Bunkyo-ku, 113-0033 Tokyo, Japan
        \and
            Instituto Nazionale di Astrofisica (INAF) - Instituto di Astrofisica e 
Planetologia Spaziali (IAPS)\\ Via Fosso del Cavaliere, 100, 00133, Rome, Italy
        \and
            Southwest Research Institute
            1050 Walnut St., Suite 300
            Boulder, colorado 80302 USA
             }

   \date{Received 19/07/2021; accepted 18/11/2021}

 
  \abstract
   {Despite the observed signs of large impacts on the surface of Ceres, there is no confirmed collisional family associated with this dwarf planet. After a dynamical and photometric study, a sample of 156 asteroids were proposed  as candidate members of a Ceres collisional family.}
   {Our main objective is to study the connection between Ceres and a total of 14 observed asteroids among the candidates sample to explore their genetic relationships with Ceres.}
   {We obtained visible spectra of these 14 asteroids using the OSIRIS spectrograph at the 10.4 m Gran Telescopio Canarias (GTC). We computed spectral slopes in two different wavelength ranges, from 0.49 to 0.80 $\mu$m and from 0.80 to 0.92 $\mu$m, to compare the values obtained with those on Ceres’ surface previously computed using the Visible and Infrared Spectrometer (VIR) instrument on board the NASA Dawn spacecraft. We also calculated the spectral slopes in the same range for ground-based observations of Ceres collected from the literature.}
   {We present the visible spectra and the taxonomy of 14 observed asteroids. We found that only two of the asteroids are spectrally compatible with Ceres’ surface. Further analysis of those two asteroids indicates that they are spectrally young and thus less likely to be members of the Ceres family.}
   {All in all, our results indicate that most of the 14 observed asteroids are not likely to belong to a Ceres collisional family. Despite two of them being spectrally compatible with the young surface of Ceres, further evaluation is needed to confirm or reject their origin from Ceres.}


   \maketitle
%

\section{Introduction}\label{sec:int}
Collisional families are thought to be the direct outcome of collisional events 
in the asteroid belt \citep{Cellino2002}. Every member in the family has similar dynamical characteristics, such as a semimajor axis, eccentricity, inclination, a longitude of perihelion, and a longitude of node, whose long-term averages are known as proper elements. Those parameters are chosen by their near constancy in collisional timing \citep{Knezevic2002}, that is to say a common collisional event with deviation by the Yarkovsky effect. However, other asteroids that do not belong to the family could accidentally acquire these orbital characteristics by a gravitational interaction with other major bodies.\\

Despite all the evidence of large impacts on Ceres' surface, no collisional family associated with the dwarf planet was found until a few years ago. \cite{Carruba} postulated the existence of this family after a dynamical study. They hypothesized that the high  gravitational influence of Ceres over the outcome and the high ejection velocity that needed to escape from it should carry the family members farther from the parent body than other collisional families associated with smaller asteroids. 
They thus looked for candidates in the so-called pristine region of the main asteroid belt. The "pristine region" is located between 2.825 and 2.960 au and is bordered by two main motion resonances that have remained almost empty of asteroids since its formation. There, they found a sample of 156 asteroids whose colors and visible albedo values 
were compatible with 
being fragments from Ceres, most of them having inclinations 
close to that of Ceres. At the end of their study, they provided a sample of 45 plausible 
candidates that satisfied their selection criteria. Further information can be found 
in \cite{Carruba}.\\

In general, dynamical studies, through the use of proper elements to identify collisional families, give rise to a number of uncertainties. Discrepancies among family members 
are found 
in
different authors' results \citep{1989aste.conf..368V,2015aste.book..297N}, 
and cosmochemical inconsistencies in some family membership appear in different 
studies
\citep{2014Icar..239...46M}. 
The
large photometric surveys 
of recent years have
provided albedos 
at
different wavelengths
that
give information 
about the composition of asteroids. This information has been combined with dynamical 
studies to 
distinguish
between family members. Thus, with a few photometric points, we 
are able to distinguish between 
the
main classes of asteroids. However, spectral properties 
in different wavelength ranges have been recognized as an invaluable tool 
in
assessing 
the real memberships of 
mutually overlapping 
families
in the space of proper elements
 \citep{Knezevic2002}.
These spectral properties enable us
to identify interlopers that are mixed within a collisional family 
and to get an idea of 
the
evolution of an asteroid. In this study, our aim is to add 
spectroscopic information to the potential Ceres family members in \cite{Carruba} 
and check if  Ceres' spectroscopic properties provide 
any
evidence,
either for
or against 
the existence of a genetic relationship.
To achieve this objective, we have observed 
14 asteroids from the list of potential family members in \cite{Carruba} and compared 
these
with Ceres' surface. The acquisition and reduction 
processes
related to the ground-based data are described in Section \ref{sec:obs}. We provide details 
on
the Ceres 
data in Section \ref{sec:dat}, 
obtained 
both
from ground-based observations 
and
by the Visible and Infrared Spectrometer (VIR) on board the NASA Dawn spacecraft. Data analysis is developed 
in Section \ref{sec:met}, and the results obtained are presented and discussed in 
Section \ref{sec:res}. 
Our final conclusions are 
expounded
in Section \ref{sec:con}.
\section{Observations and data reduction}\label{sec:obs}
The sample of observed asteroids 
was
selected from the dynamical study of \cite{Carruba}, 
who
identified a total of 156 asteroids as 
potential 
members of the 
Ceres
collisional family, which have color photometry and/or albedos compatible 
with those of C-type asteroids, a diameter lower than 20 km, and a
location
in the pristine region. 
The authors mark those asteroids with a proper 
inclination in the range 0.098 < $\sin(i)$ < 0.239, which is the expected range
for
the Ceres family. Finally, they selected 45 asteroids 
that
satisfied the more restrictive criteria they imposed with the aim of 
reducing contamination from the halo of large local families 
such as those of
Charis, Eos, and Koronis 
(for further information, see \citealt{Carruba}). To simplify the notation, we 
label the asteroids among the more restrictive sample as subsample a, the asteroids 
with compatible proper inclination as subsample b, and the most general sample as subsample 
c. We 
were
able to observe a total of 14 asteroids from the whole sample: 
five
from subsample a, 
three
from subsample b, and 
six
 asteroids from the most general subsample c.  

\begin{table*}[!ht]
\centering
\begin{tabular}{llccccccccll} 
    \hline
    Asteroid & Date & UT & T$_{\rm exp}$ & seeing & airmass & m$_V$ & $\alpha$     & $\Delta$ & $r$ & Solar & sample \\
             &      &    & (s)       & ('')   &         &       & ($^{\circ}$) & (au)     & (au)& analogs  &        \\          
    \hline\hline
    61674  & 12/04/2017    & 22:16 & 3x300 & 1.4 & 1.15 & 18.1 & 15.5   & 1.74 & 2.56 & 1,3 & a \\
    66648  & 10/04/2017    & 01:16 & 3x600 & 1.8 & 1.13 & 19.8 &  6.0   & 2.32 & 2.28 & 2,1  & a \\
    20094  & 10/04/2017    & 05:14 & 3x600 & 1.8 & 1.20 & 19.7 & 14.4   & 2.50 & 3.21 & 2,1 & a \\
    222080 & 10/04/2017    & 03:02 & 9x300 & 1.7 & 1.20 & 20.1 &  3.4   & 2.30 & 2.29 & 2,1 & a \\
           & 17/04/2017    & 02:17 & 3x900 & 0.8 & 1.20 & 20.1 &  3.2   & 2.29 & 3.28 & 2,1 &   \\
    261489 & 17/04/2017    & 01:06 & 3x900 & 0.8 & 1.40 & 21.1 & 15.3   & 2.63 & 3.27 & 2,1 & a \\
    \hline
    5994   & 10/04/2017    & 00:15 & 3x300 & 1.8 & 1.20 & 18.0 & 15.0  & 2.95 & 3.50 & 2,1 & b \\
    23000  & 10/04/2017    & 02:00 & 3x300 & 2.3 & 1.07 & 18.1 &  7.6   & 2.02 & 2.97 & 2,1 & b \\
    198403 & 10/04/2017    & 04:16 & 3x300 & 1.8 & 1.24 & 20.3 & 20.5   & 2.25 & 3.18 & 2,1 & b \\
           &               & 04:32 & 4x450 & 1.8 & 1.30 &      &        &      &      &                      &   \\
    \hline
    6671   & 12/04/2017    & 21:23 & 3x300 & 1.5 & 1.14 & 18.5 & 18.0   & 2.86 & 3.19 & 1,3 & c \\
    22540  & 12/04/2017    & 21:51 & 3x300 & 1.4 & 1.05 & 19.1 & 19.1   & 2.21 & 2.78 & 1,3 & c \\
    20095  & 17/04/2017    & 02:52 & 3x300 & 1.1 & 1.50 & 19.0 &  5.0   & 2.08 & 3.06 & 2,1 & c \\
    121281 & 17/04/2017    & 01:25 & 3x600 & 1.1 & 1.40 & 19.6 &  6.2   & 2.01 & 2.98 & 2,1 & c \\
    38466  & 17/04/2017    & 03:21 & 3x300 & 1.4 & 1.60 & 19.0 & 10.5   & 1.90 & 2.81 & 2,1 & c \\
    155547 & 17/04/2017    & 02:07 & 3x600 & 1.1 & 1.30 & 19.9 &  1.7   & 2.25 & 3.25 & 2,1 & c \\
    \hline\\
\end{tabular}
\caption{Observational details of the observed asteroids. The solar analog stars used are 1:SA102-1081, 
2:SA107-998, and 3:SAM67-1194. The last column refers to the subsamples identified in this work from 
the general list of potential  Ceres family
members in \cite{Carruba}.}
\label{t:Sample}
\end{table*}

Asteroid spectra were obtained through ground-based observations carried out with the OSIRIS instrument \citep{2000SPIE.4008..623C,2010ASSP...14...15C} installed on the 10.4 m Gran Telescopio Canarias (GTC) operated by the Instituto de Astrof\'isica de Canarias at Roque de los Muchachos Observatory (ORM) in La Palma. Observations were executed under the "filler" (Band-C) program GTC68-17A. Filler programs are aimed to exploit telescope schedule gaps during the night that cannot be used for proposals that require better seeing and weather conditions. Observational details of the spectra obtained are shown in Table \ref{t:Sample}, including the asteroid number, date, UT starting time, exposure time, seeing and airmass values, apparent visual magnitude ($m_V$), phase angle ($\alpha$), and distances to the Earth ($\Delta$) and to the Sun ($r$) at the time of observing. Table \ref{t:Sample} shows that seeing values varied from night to night, or even during the same night. This explains the differences in signal-to-noise ratios (S/Ns) at different asteroid spectra.

We used the R300R grism, which has a dispersion of 7.74 \AA/pixel for a $0.6^{\prime\prime}$ slit and covers a wavelength range between 0.48 and 1 $\mu$m. A second order cutting filter is used with the grism that discards wavelengths  beyond 0.92 $\mu$m. We used a 
$2.52^{\prime\prime}$ 
slit width to account 
for variable seeing conditions. The slit was oriented to the parallactic angle to minimize loses due to atmospheric dispersion and to maximize the flux. For each asteroid, we obtained at least three spectra,  offsetting the object $10^{\prime\prime}$ in the slit direction between  
individual observations. To obtain 
reflectance spectra
of the asteroids, we observed two solar 
analog stars on each night at a similar airmass to that of the targets. 
The stars used are 
specified in Table \ref{t:Sample}.

The data were reduced 
using both IRAF\footnote{Image Reduction and Analysis Facility: 
\href{http://iraf.noao.edu}{http://iraf.noao.edu}} and a 
pipeline 
developed
in Python 2.7. 
The data reduction steps included standard bias subtraction and flat field correction. Using 
the
\textit{apall} task in 
IRAF, we extracted 1D spectra from the 2D images, selecting an extraction aperture and a 
background region that changed for each target. For asteroids having low S/N 
(222080, 261489, and 198403),
we aligned and summed the individual 2D spectra to 
increase the S/N before 
extraction. Wavelength calibration was 
performed
using Hg-Ar, 
Ne, and Xe lamps. The spectra for each target were averaged, and the resulting spectrum was divided by the spectra of the two solar analogs observed each night and then normalized to unity at 0.55 $\mu$m. To obtain a reflectance spectrum of each asteroid, we averaged those two spectra obtained from two solar analogs. As a final step, we applied a phase correction to the spectra to the standard viewing geometry phase angle of 30$^{\circ}$, following the procedure described in \cite{2017A&A...598A.130C}, who proposed the following relation between the slope in the visible and phase angle: $S_{\rm VIS}=-9.4\times10^{-7}\alpha + 1.1\times10^{-2}$, with the angular coefficient expressed in [k$\AA^{-1}\deg^{-1}$] and the intercept in [k\AA$^{-1}$], based on the observations of Ceres. The spectra obtained are shown in Fig. \ref{f:Res:asteroid_spectra} in 
red. 
The error bars correspond to the difference between the two spectra obtained (owing to the use of two different solar analogs). For targets with a low S/N, we binned the  spectra to a total of 50 points, corresponding to a binning box of about 100 $\AA$.

\section{Ceres data}\label{sec:dat}
To compare our visible spectra with the spectra for Ceres, we searched the literature for any spectral data already published from ground-based and space-based data. We compiled different ground-based visible spectra of Ceres from previous studies: one in the S$^3$OS$^2$ survey \citep{2004Icar..172..179L}, with a phase angle of 16$^{\circ}$; one in  the SMASSII survey \citep{2002Icar..158..106B}, with a phase angle of  18.5$^{\circ}$; and another obtained by \cite{1992Icar..100...85V} at a phase angle of 6.1$^{\circ}$. We also included the six spectrophotometric observations from the Eight color Asteroid Survey (ECAS) \citep{1985Icar...61..355Z} obtained with phase angles between 8$^{\circ}$ and 22$^{\circ}$. We excluded the spectra from the 24 color Asteroid Survey \citep{1979aste.book.1064C,1984Icar...59...25M} because of their high noise at the longer wavelengths of the visible range. We also applied a phase correction to the ground-based data for Ceres, using the same procedure as in \cite{2017A&A...598A.130C}.
All these spectra have a wavelength range comparable to or greater than that of our spectra obtained with the OSIRIS instrument. For the ECAS data, we computed the reflectance values at OSIRIS wavelength limits (0.49 and 0.92 $\mu$m) by interpolating the nearest filters.
The area covered by the different spectra is represented by a gray hatch in the upper left panel of Figure \ref{f:Res:asteroid_spectra}. We computed the mean of all the ground-based spectra in the same wavelength range as that of our observations and used this ground-based mean spectrum to compute the spectral slopes. Again, the errors are given by the variations between the different spectra.\\

Regarding the space-based data, the NASA Dawn spacecraft visited Ceres in 2015. From the first observations, we know that there is spectral variation on its surface \citep{2016P&SS..134..122N}. \cite{2015Icar..260..332R} found variations in the band depth near 1.1 $\mu$m that are to be correlated with albedo, and in the spectral slope related to the phase angle and longitude. If our observed asteroids did indeed originate from a collision with Ceres in the past, they would show a spectral variation similar to that observed on the surface of the dwarf planet. In order to compare the asteroid spectra with the observed spectral variation on the surface of Ceres. We used the spectral slope data calculated by \cite{Rousseau} with the visible channel of the VIR spectrometer \citep{2011SSRv..163..329D}. VIR is a mapping spectrometer with a moderate spectral resolution (0.002 $\mu$m in the visible range) and it covers a wider wavelength range than OSIRIS (from 0.25 $\mu$m to 1.0 $\mu$m). The mapping capability of VIR allowed us to provide almost complete coverage of the surface of Ceres during the Dawn mission.

\cite{Rousseau} produced spectral parameter maps of the surface of Ceres for latitudes from 60$^{\circ}$S to 75$^{\circ}$N. They calculated the spectral slope for the whole surface in three wavelength ranges: the visible to near-ultraviolet ($S_{VNUV}$, 0.405-0.465 $\mu$m), the visible  ($S_{VIS}$, 0.480-0.800 $\mu$m), and the visible to near-infrared ($S_{VNIR}$, 0.800-0.950 $\mu$m). They also calculated the albedo at 0.550 $\mu$m and several color ratios. In this study we used those spectral maps only at medium latitudes (from $-60^{\circ}$ to 60$^{\circ}$) to avoid high phase angles, shadows, and bright regions that might otherwise be introduced by high incidence and emission angles. In order to adapt the map's spatial resolution to the size of our sample asteroid ($\sim10\;\mbox{km}$), we binned it in $1^{\circ}\times1^{\circ}$ bins, $\sim100\;\mbox{km}^2$ at medium latitudes of the surface of Ceres.

\section{Data analysis}\label{sec:met}

After obtaining the final reflectance spectra for each ground-based observed asteroid, we classified them using the so-called Bus taxonomy \citep{2002Icar..158..146B}, which is one of the most comprehensive taxonomies based on a total of 1447 visible spectra of asteroids. We carried out a visual classification following the criteria described in \cite{2002Icar..158..146B}, which include the following: the existence of local maximum, a wavelength at which the slope changes, how much slope changes before and after the maximum (if it exists), and a visible slope value. We also used a computational classification taking advantage of the online tool M4AST.\footnote{\href{http://spectre.imcce.fr/m4ast/index.php/index/home}{http://spectre.imcce.fr/m4ast/index.php/index/home}} \citep{m4ast} This tool fits a curve to the data and compares it with taxonomic classes defined by \cite{2009Icar..202..160D}, which is an extension to the near-infrared of the Bus taxonomy, using $\chi^2$ to evaluate this comparison. In terms of the Bus taxonomy, Ceres belongs to the primitive C complex, which is characterized by a flat, featureless spectra.

An absorption band at 0.7 $\mu$m is usually found in primitive asteroids. The band is associated with Fe bearing hydrated minerals, such as phyllosilicates \citep{1989Sci...246..790V}, implying the presence of liquid water at some instances in the the lifetime of the asteroid  \citep{1994Icar..111..456V,1999A&AS..135...65F,2012Icar..221..744R,2014Icar..233..163F}. This band is also interesting because it is related to the band observed at 3 $\mu$m, which is also associated with hydrated silicates and present in Ceres \citep{1994Icar..111..456V}. When the 0.7 $\mu$m is present, we usually observe the 3 $\mu$m band. However, the converse is not true; that is to say,\ the absence of an absorption band at 0.7 $\mu$m does not imply the absence of hydrated silicates \citep{1994Icar..111..456V}. When the 0.7 $\mu$m band was found to be present in our asteroid spectra, we studied it by subtracting the continuum (a line connecting the two reflectance maxima in the wings of the band) and fitting a Gaussian function. From this fit we obtained the central wavelength (the minimum of the Gaussian) and the depth of the band (as a percentage). The error in these parameters is given by the differences between the resulting spectra when dividing the solar analogs.

To know how different an asteroid could be from Ceres, we first studied the distribution of the spectral slope for the surface of Ceres. To do this, we used the global maps with calculated spectral slopes in \cite{Rousseau} and computed the spectral slopes in our target asteroids using the same definition as in \cite{Rousseau} and adopted from \citealt{2015A&A...583A..31C, 2017A&A...598A.130C}. The spectral slope in a wavelength range $\left(\lambda_1, \lambda_2\right)$ is given by

\begin{equation} S_{\left(\lambda_1, \lambda_2\right)} =  \frac{\left( \frac{I}{F} \right)_{\lambda_2} - \left( \frac{I}{F} \right)_{\lambda_1}}{\left( \frac{I}{F} \right)_{\lambda_1}\times\left(\lambda_2 - \lambda_1\right)}, 
\label{ec:Sam:slope}
\end{equation}
\noindent 

where ($I/F$)$_{\lambda_1}$ is the calibrated radiance factor or reflectance measured at $\lambda_1$. With this definition, \cite{Rousseau} computed three different slopes, $S_{\rm VNUV}$, $S_{\rm VIS}$, and $S_{\rm VNIR}$, as described in Section \ref{sec:dat}. The instrumental setup used with OSIRIS at the GTC provides an effective wavelength range between 0.49 and 0.92 $\mu$m; we thus computed the asteroid's visible spectral slope between 0.49 and 0.80 $\mu$m and the visible near-infrared slope between 0.80 and 0.92 $\mu$m. As those spectral regions are mainly flat and linear in C complex asteroids, except for the 0.7$\mu$m band (if present), the difference when using a slightly shorter wavelength range is not significant.

\section{Results and discussion}\label{sec:res}

\subsection{Comparison between the sample and Ceres}\label{sec:Res:sub:comparison}

\begin{figure*}[!ht]
\centering
\includegraphics[width=\textwidth]{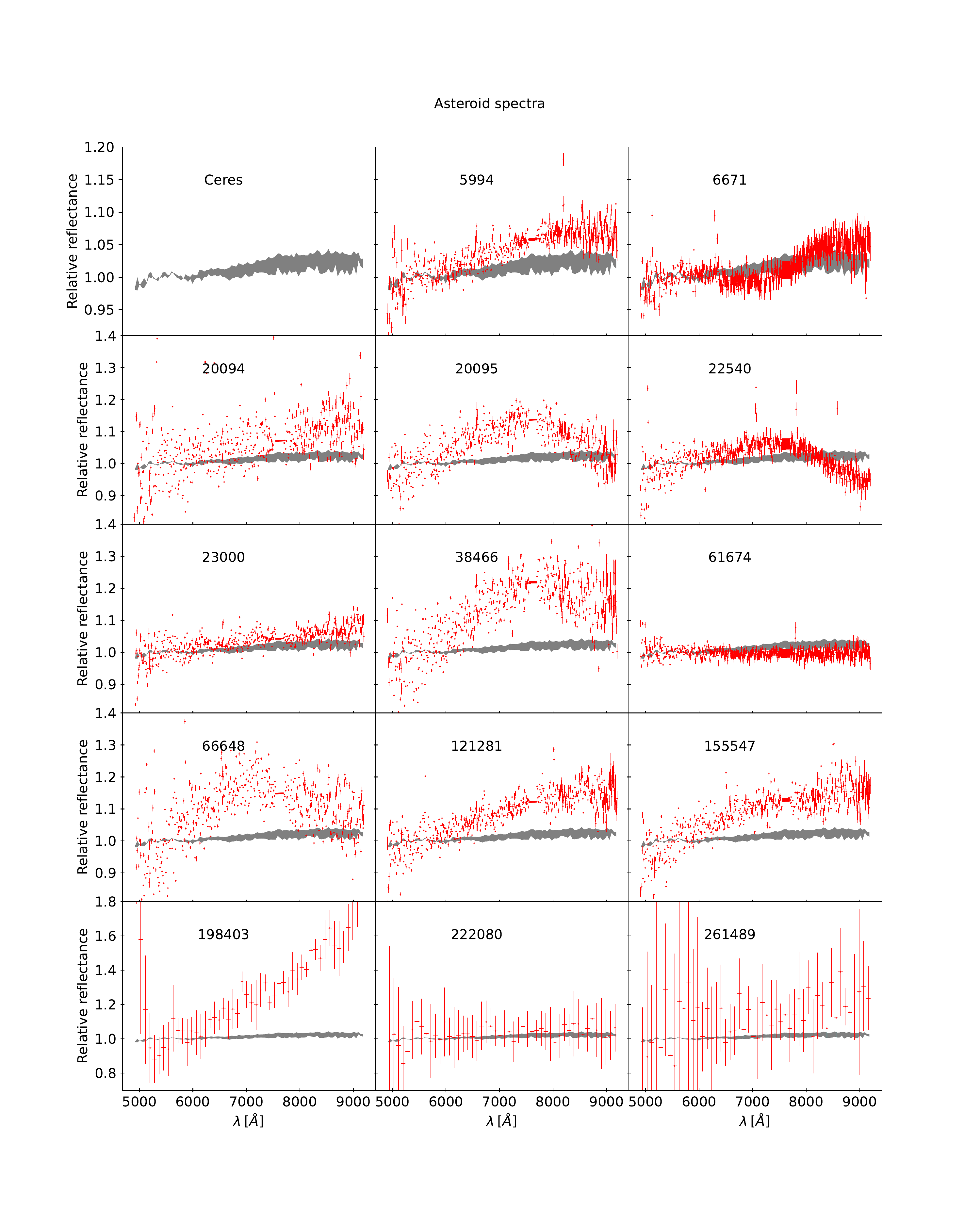}
\caption{Visible reflectance spectra of the observed asteroids (in red), normalized to unity at 0.55 $\mu$m and phase corrected. The upper left panel shows the area covered by the different ground-based spectra of Ceres used in this study (see Section \ref{sec:dat} for details). This variation is superposed on the spectra of each asteroid in the other panels as a gray hatched area. For asteroids with a low S/N, we applied a binning factor of $\sim$10, obtaining 50 point spectra.}
\label{f:Res:asteroid_spectra}
\end{figure*}

The visible spectra of the observed asteroids are shown in Figure \ref{f:Res:asteroid_spectra}. The variation in ground-based spectra of Ceres used in this study is shown as a gray hatched area. This variation can be explained by several observational issues, such as bad slit centring, solar analog discrepancies, and differences in airmass between the asteroid and the solar analogs. As the spectra were collected from the literature, we do not have access to such information, with the exception of the phase angles, for which we already applied a correction following \cite{2017A&A...598A.130C}.

We observed a variety of asteroid classes in our sample. As Ceres is a carbonaceous asteroid, one would expect the asteroids of its family to also be carbonaceous. Taxonomic classes with a carbonaceous composition belong to the C complex, and X types with a low albedo (mainly Xc types). Other primitive taxonomic classes such as end member T and D types have steep, red slopes associated with the presence of processed organics typical of outer belt asteroids, Trojans, and cometary nuclei which are supposed to form in the cold outer solar system \citep{2019NatAs...3..910F} and, therefore, to be less compatible with an origin in a C-type asteroid as Ceres. We show the taxonomic classification and computed slopes in Table \ref{t:TaxAst}. The results of our visual classification (following the procedure described in Section \ref{sec:met}) are listed in the second column of the table, while the third and fourth columns show the best two fits (with their associated $\chi^2$ value) obtained with the M4AST tool. As can be seen in the table, there are some discrepancies in the results between the visual and the computational classifications in terms of the specific taxonomic classes, but the agreement is generally quite good. Considering these results, we assigned the asteroids to one of the three major taxonomic complexes (the C, X, or S complex), as can be seen in the fifth column of the table.

Regarding the major discrepancies found, the M4AST tool classifies asteroid 22540 as a C-type object. However, the observed maximum at 0.75 $\mu$m and the drop in reflectance upwards of 1 $\mu$m suggest an S-type classification after visual inspection of its spectrum, which is even flatter than usual S-type asteroids. In the case of the three asteroids having low S/N (222080, 198403, and 261489), we used the general behavior of their spectra to classify them. Asteroid 222080 has a flat and featureless spectrum that could belong to spectral types C or X, but the high VNIR spectral slope made us include it in the X complex. The high spectral slope of asteroid 198403 suggests that it is a D-type asteroid. The dispersion in the data points of the 261489 spectrum makes it almost impossible to provide a reliable classification. The variety of taxonomic complexes and classes that we have found in our sample of 14 asteroids (C, X, S, and D) indicates that not all of them belong to the same collisional family. Four out of 14 of our observed asteroids belong to the C complex: 61674 from subsample a; 5994 and 23000 both have an inclination in the same range as Ceres; and 6671 from subsample c (the initial sample in \citealt{Carruba}). We found another four asteroids in the X complex: 20094 and 222080, from subsample a, and 121281 and 155547 from subsample c.

\begin{table*}[t]
\centering
\begin{tabular}{llllcccl} 
    \hline
    Asteroid  & Visual & \multicolumn{2}{c}{M4AST - $\chi^2$ (10$^{-3}$)} & Complex 
& $S_{\rm VIS}$ ($10^{-5}\AA^{-1}$) & $S_{\rm VNIR}$ ($10^{-5}\AA^{-1}$) & Sample\\
    \hline\hline
    61674  & B      & Ch  - 0.182  & B   - 0.342 & C & -1.7$\pm$0.6 & -0.2$\pm$0.6 & a\\
    6671   & Ch     & Cb  - 0.191  & Cgh - 0.274 & C & 1.4$\pm$0.7 & 1.6$\pm$0.5 & c\\
    5994   & C      & Cgh - 0.192  & C   - 0.235 & C & 4.4$\pm$0.3 & -1.0$\pm$0.7 & b\\
    23000  & C/Cb   & Cgh - 0.077  & C   - 0.138 & C & 3.4$\pm$0.8 & 1.6$\pm$0.6 & b\\
    20094  & X      & Xc  - 0.169  & X   - 0.224 & X & 7.9$\pm$2.0 & 0.9$\pm$0.7 & a\\
    121281 & X      & X   - 0.242  & Xc  - 0.253 & X & 5.6$\pm$0.8 & 0.9$\pm$2.2 & c\\
    155547 & Xk     & Xe  - 0.100  & Xk  - 0.455 & X & 7.3$\pm$1.0 & 1.3$\pm$2.2 & c\\
    222080 & C/X?   & O   - 15.773 & V   - 17.429 & X & 2$\pm$4 & 6.3$\pm$1.3 & a\\
    198403 & D?     & Cb  - 31.374 & X   - 32.408 & D & 16.0$\pm$0.3 & 40.3$\pm$1.3 & b\\
    66648  & S      & Sq  - 0.448  & Sr  - 0.609 & S & 13$\pm$6 & -4.2$\pm$1.4 & a\\
    20095  & Sq     & Sq  - 0.458  & Sr  - 0.654 & S & 5.4$\pm$0.8 & -9.0$\pm$2.0 & c\\
    38466  & S      & S   - 0.353  & Sv  - 0.393 & S & 8.2$\pm$0.8 & -7.0$\pm$2.0 & c\\
    22540  & Sq     & B   - 0.916  & Cg  - 1.166 & S & 3.7$\pm$0.9 & -8.3$\pm$0.5 & c\\
    261489 & -      & A   - 21.854 & L   - 26.925 & - & 1$\pm$4 & 8$\pm$4 & a\\
    
    \hline
    Ceres  & C      & C   - 0.0096 & Cb  - 0.1087 & C & 1.2$\pm$0.5 & 0.2$\pm$0.9\\
    \hline\\
\end{tabular}
\caption{Taxonomic classification (both visual and using the M4AST online tool), as well as the computed spectral slopes for the observed asteroids. The last column shows the sample to which asteroids belong (see main text for details).}
\label{t:TaxAst}
\end{table*}

\begin{figure}[h]
\centering
\includegraphics[width=\columnwidth]{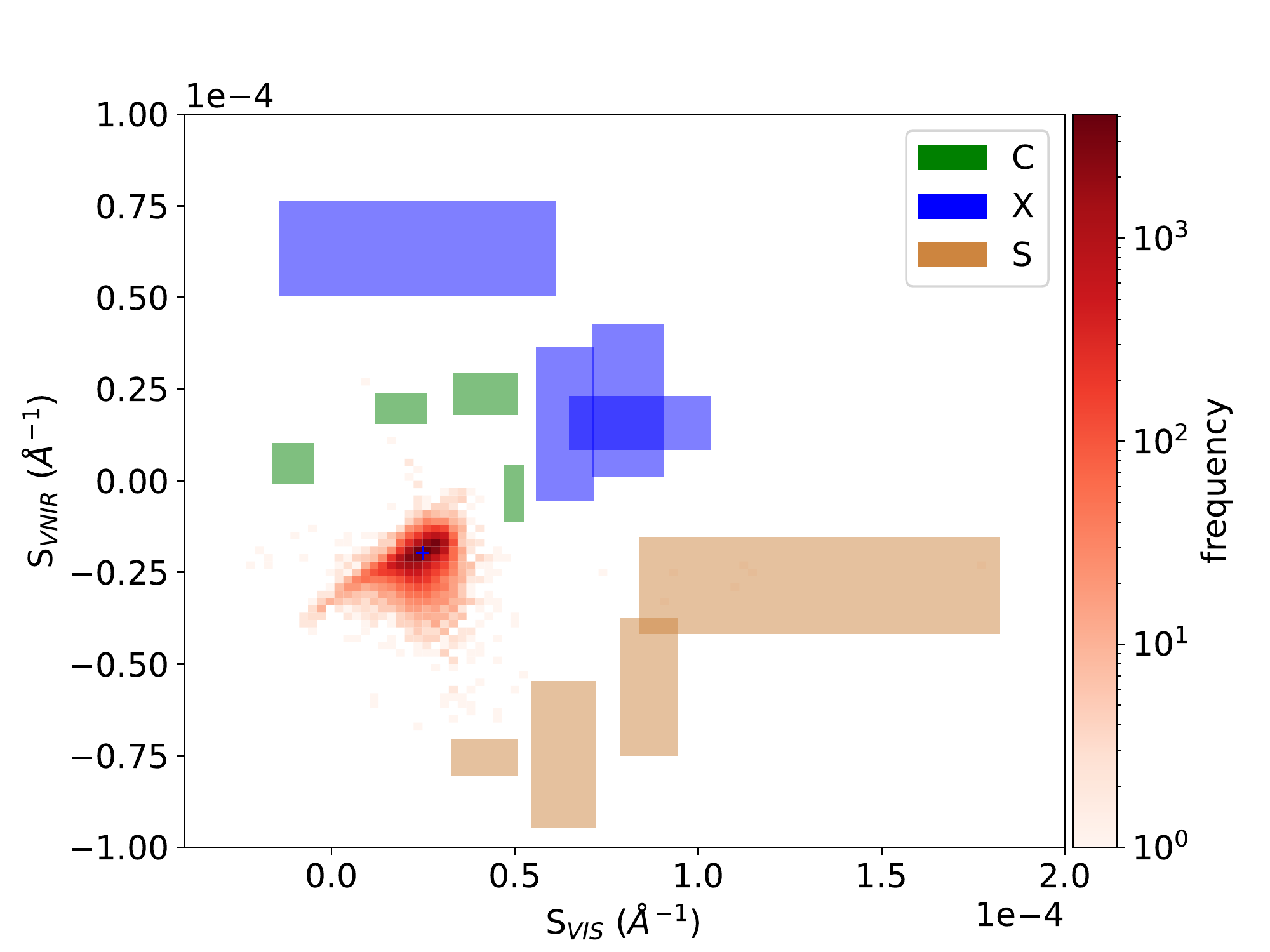}
\caption{2D density distribution of spectral slopes $S_{VIS}$ and $S_{VNIR}$ in the surface of Ceres, computed from the Dawn/VIR data, provided by \cite{Rousseau} and binned in $1^{\circ} \times1^{\circ} $ boxes. Computed slopes (and their errors) for the targeted asteroids belonging to the C complex (green boxes), X complex (blue boxes), and S complex (yellow boxes). The median slope values of the surface of Ceres are represented by a blue cross.}
\label{f:Res:2d_gen_hist}
\end{figure}

\begin{figure}[h]
\centering
\includegraphics[width=\columnwidth]{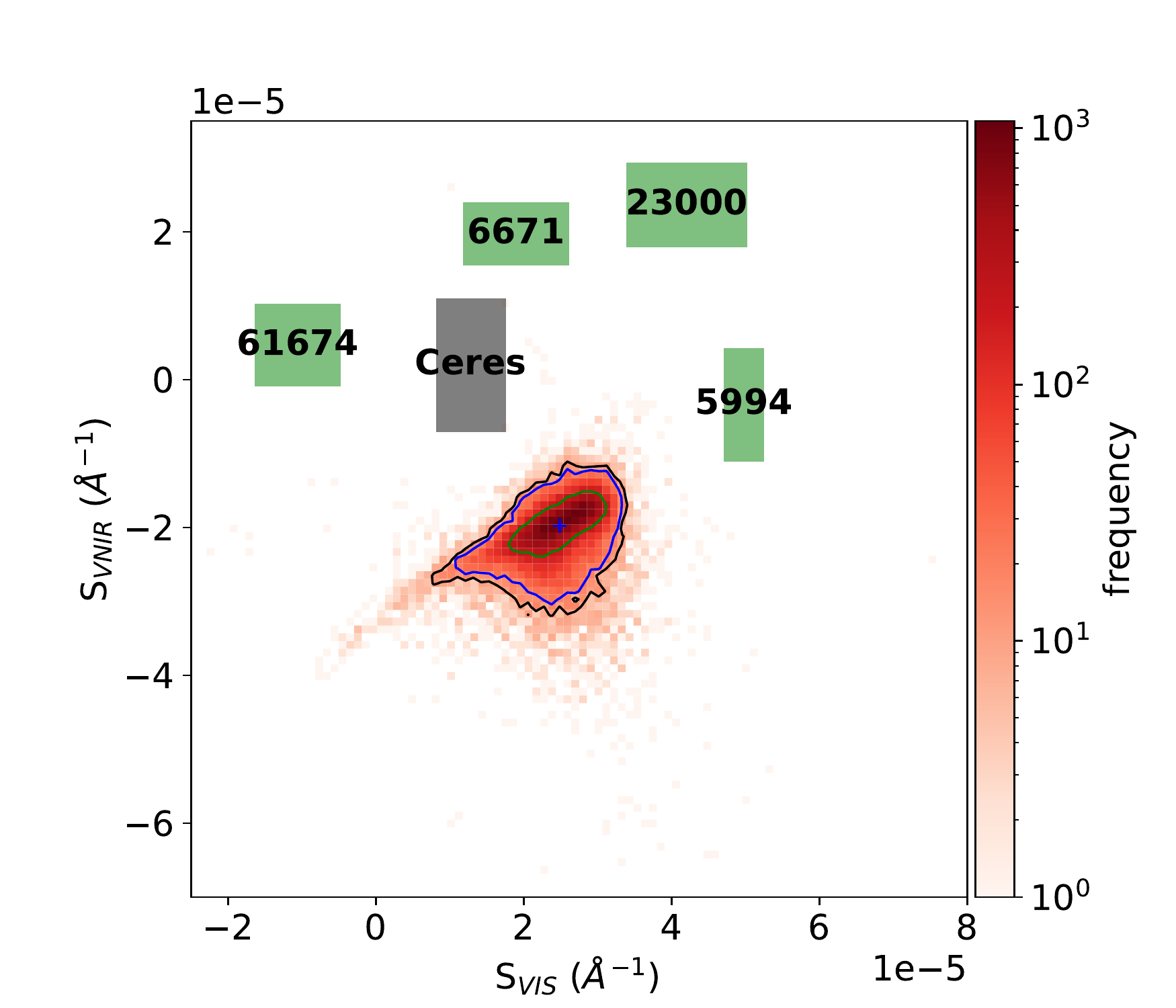}
\caption{Same as Fig. \ref{f:Res:2d_gen_hist}, but with a zoom into the densest region. Computed slopes (and their errors) for both the targeted C complex asteroids and ground-based average spectrum of Ceres are overplotted with green and dark gray boxes, respectively. The median slope value of the surface of Ceres are represented by a blue cross. The black, blue, and green contours correspond to the 99.7\%, 95\%, and 68\% most usual values, i.e., 3$\sigma$, 2$\sigma$, and 1$\sigma$ distance to the median.}
\label{f:Res:2d_C_hist}
\end{figure}

To go further, we analyzed the slope distribution on the surface of Ceres, using the spectral data provided by \cite{Rousseau} and described in Section \ref{sec:dat}. The distribution of slopes on the surface of Ceres, binned in $1^{\circ} \times1^{\circ}$ areas, is shown in the ($S_{\rm VIS}$, $S_{\rm VNIR}$) density plot of Figure \ref{f:Res:2d_gen_hist}. Where we have superimposed the slopes (and their errors) computed for the asteroids belonging to the C complex as green boxes, those belonging to the X complex as blue boxes, and those belonging to the S complex as light brown boxes. It is interesting to note here that those asteroids in the C complex are actually those located closer to the populated region in this density plot. Figure \ref{f:Res:2d_C_hist} is a zoom into this densest region; we have also plotted in dark gray the computed slopes for the average ground-based spectrum of Ceres (see Table \ref{t:TaxAst}). The distribution shows that the median slope values (blue cross in Fig. \ref{f:Res:2d_C_hist}) are $S_{\rm VIS} = 2.49 \times 10^{-5}$ $\AA^{-1}$ and $S_{\rm VNIR}$ = $-1.94\times 10^{-5}$ $\AA^{-1}$. We notice an asymmetry in the distribution in the bottom left side from the median point, where there seems to exist an over dense direction. This is studied in Section \ref{sec:Res:sub:Privileged_direction}.

\begin{figure}[h]
\centering
\includegraphics[width=\columnwidth]{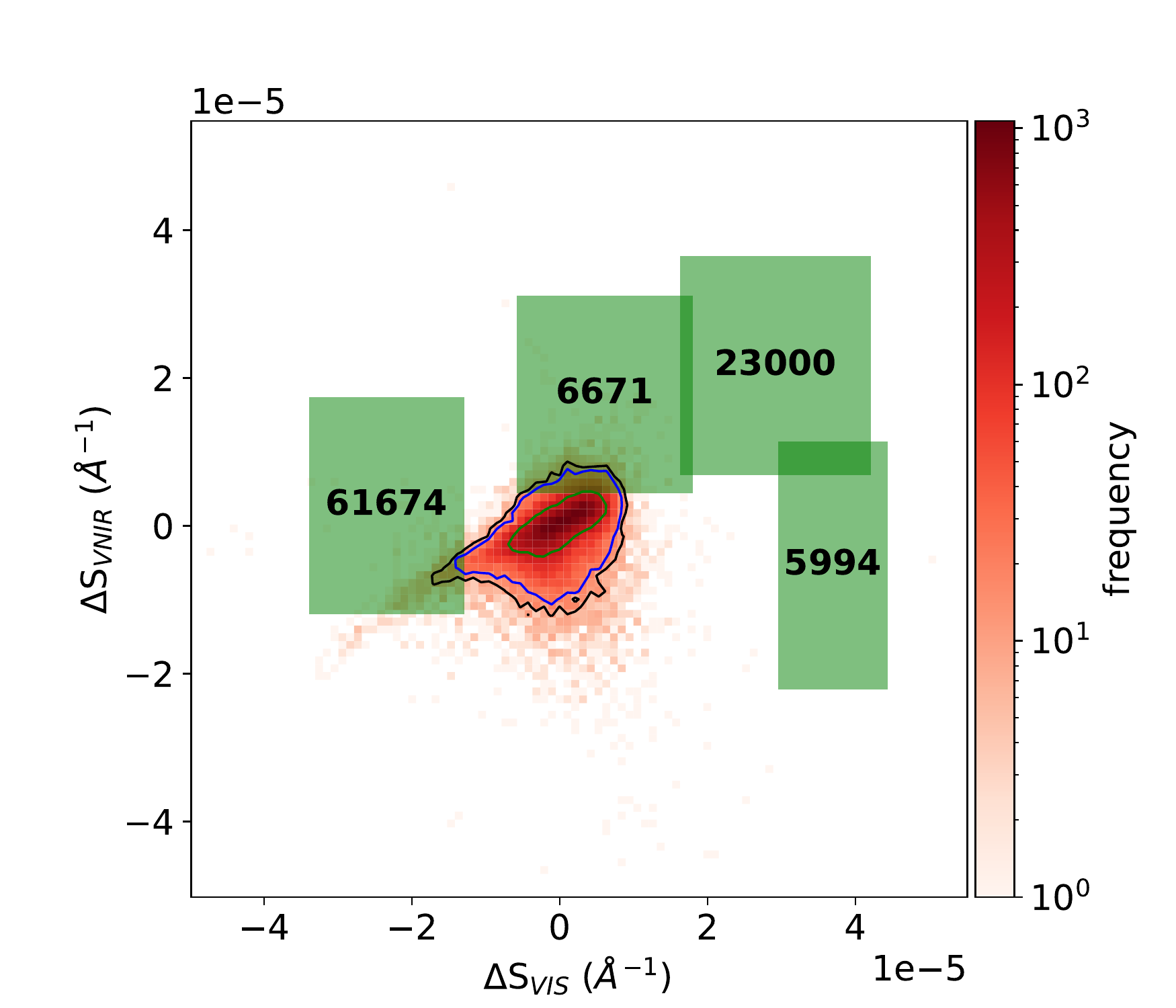}
\caption{ Same as Fig. \ref{f:Res:2d_C_hist}, but for the differences in $S_{VIS}$ and $S_{VNIR}$ between the C-type asteroids and ground-based spectrum of Ceres. The distribution of slopes on the surface of Ceres, as measured by Dawn/VIR, is relative to the median value.}
\label{f:Res:2d_hist_shifted}
\end{figure}

As we can see in Fig. \ref{f:Res:2d_C_hist}, there is a difference between the slopes computed from ground-based spectra of Ceres (dark gray box) and the values obtained from spacecraft data. According to \cite{2016RScI...87l4501C}, Dawn/VIR spectra are affected by a positive slope in the VIS to NIR range when compared to ground-based spectra of the same target, an effect for which the origin is not currently understood. In addition, several effects (see first paragraph of Section \ref{sec:Res:sub:comparison}) can affect the spectral slopes when measured from the ground. Thus, with the aim of avoiding all of these discrepancies, we only compared the slope distributions in a relative way: asteroids from the ground relative to Ceres from the ground and slopes in the surface of Ceres relative to their median value. To visualize the results more clearly, we provide the results of Fig. \ref{f:Res:2d_C_hist} in Fig. \ref{f:Res:2d_hist_shifted} in this relative way. The black, blue, and green contours correspond to the 99.7\%, 95\%, and 68\% most usual values, that is the 3$\sigma$, 2$\sigma$, and 1$\sigma$ distance to the median, respectively. In other words, if the distance from the asteroid spectral slopes to the center of the distribution in this new figure exceeds the 1, 2, or 3$\sigma$ contours, it means that the slope values of the asteroid are further from those of Ceres than 68\%, 95\%, or 99.7\% of the slope values of the surface of Ceres to the median value. Thus the asteroid is less likely to be constituted of the same material as the surface of Ceres. Based on this analysis, we may conclude that, taking the errors into account, the slopes of 6671 lies inside the 1$\sigma$ contour and the slopes of 61674 have a difference with the slopes of Ceres smaller than the 2$\sigma$ contour. Furthermore, 6671 has an absorption band at 0.7 $\mu$m that we study further in the next section. Another hint is that asteroid 61674, which belongs to the preferred samples of \cite{Carruba}, lies near the overdense direction observed in the histogram and mentioned earlier in this section. Asteroids 5994 and 23000 are further from the 3$\sigma$ contour, meaning that its slope values differ by more than the 997\% variation of the surface of Ceres. That leaves two out of the 14 initial asteroids as candidates to be spectrally similar candidates to Ceres. 

\subsection{Asteroid 6671: A 0.7$\mu$m absorption band}\label{sec:Res:sub:absorption_band}

\begin{figure}[!ht]
\centering
\includegraphics[width=\columnwidth]{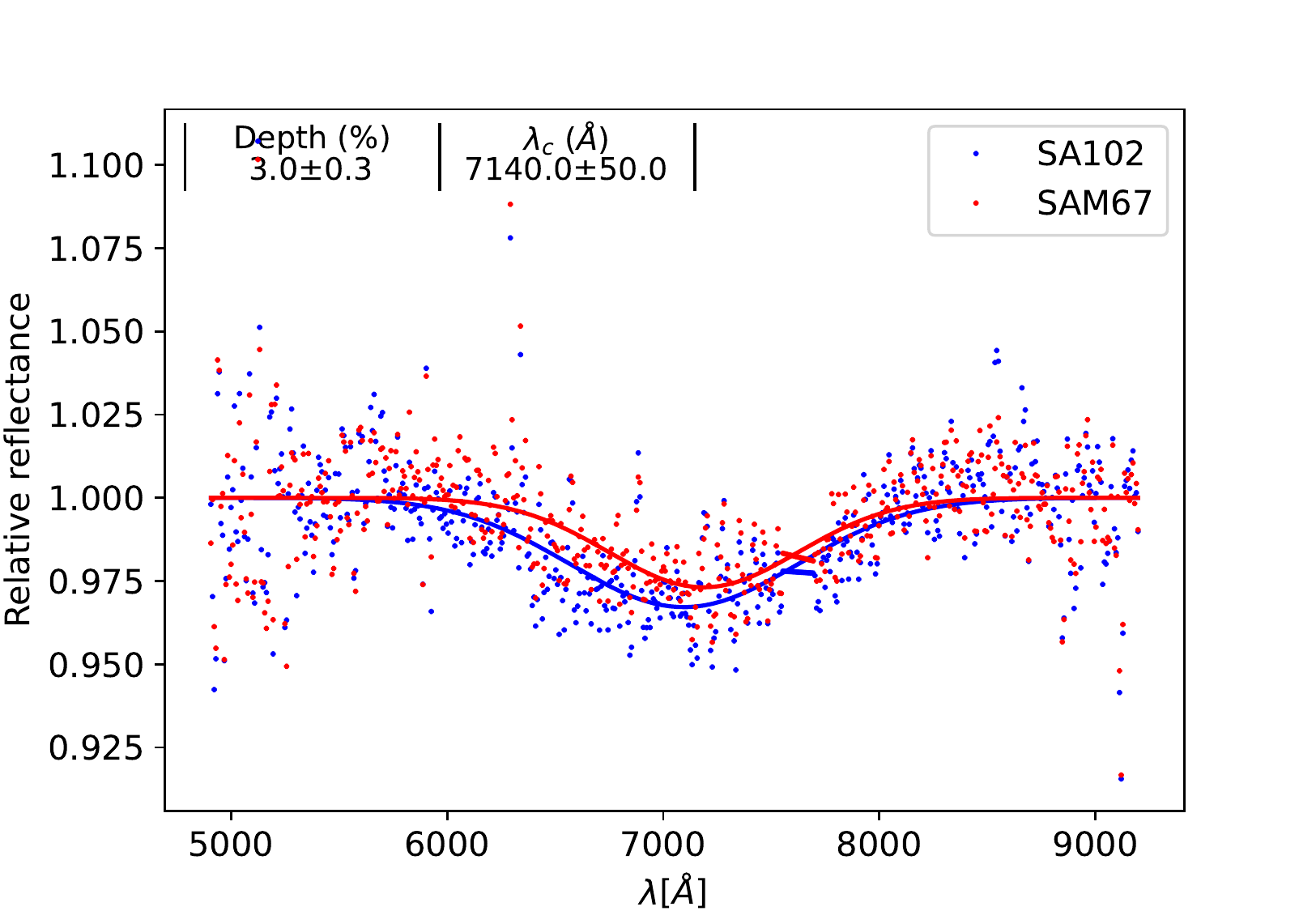}
\caption{Individual spectra of asteroid 6671 obtained using two solar analogs (Landolt SA102-1333 - SA102, and SAM67-1194 - SAM67) and with the continuum removed. Gaussian fits to the band for both spectra are represented with continuous lines. The average band depth (\%) and wavelength position of the center of the band ($\lambda_c$) are also shown.}
\label{f:Res:hydration_band}
\end{figure}

As we can see in Fig. \ref{f:Res:asteroid_spectra}, asteroid 6671 has an absorption band around 0.7 $\mu$m. The two individual spectra of 6671 obtained using two different solar analogs are shown in Fig. \ref{f:Res:hydration_band} after continuum removal (following the  procedure described in Section \ref{sec:met}), together with the Gaussian fit and the corresponding derived parameters (wavelength position of the center of the band and band depth, as percentages). 

The detection of a 0.7 $\mu$m in ground-based spectra of Ceres has only been reported by \cite{2015A&A...575L...1P} with an absorption depth smaller than 0.7\%. On the contrary, absorption bands in the 3 $\mu$m region are clearly detected, both in ground-based and space data, and they are associated with the presence of phyllosilicates, iron-rich clays, and carbonates \citep{1981Icar...48..453L,2005A&A...436.1113V,2006Icar..185..563R,2015Natur.528..241D,2019PASJ...71....1U}. In addition, a subtle absorption band at 0.6 $\mu$m was reported in Ceres by \cite{Vilas1993} and later confirmed by \cite{1999A&AS..135...65F}, which was produced by charge transfer in aqueous alteration products. Interestingly, \cite{2019Icar..328...69R} identified an absorption band at 0.7$\mu$m around the Occator crater of Ceres using data from the Dawn Framing Camera, and they measured the band depth and the wavelength position of its center: depth $ = 3.4\pm 1.0$\% and $\lambda_c=6980\pm 70$ \AA. Our results for asteroid 6671 (depth $=3.0\pm 0.3$\% and $\lambda_c=7140\pm 50$ \AA) are in good agreement with those obtained by \cite{2019Icar..328...69R}. It is important to note here, however, that these similarities are not unique or exceptional: the values obtained for the 0.7 $\mu$m absorption band in 6671 are also in agreement with typical values found in other primitive asteroids throughout the main belt (2.8$\pm$1.2\% and 6914$\pm$148 \AA, \citealt{2014Icar..233..163F}) and in the primitive collisional families of the inner belt  (2.8$\pm$1.3\% and 7065$\pm$160 \AA \, for Erigone in \citealt{2016A&A...586A.129M}; 2.2$\pm$0.6\% and 7000$\pm$150 \AA \, for Klio, 3.1$\pm$0.8\% and 7110$\pm$130 \AA \, for Chaldaea, and 2.9$\pm$1.2\% and 7100$\pm$150 \AA \, for Chimaera in \citealt{2018A&A...610A..25M}). 

Furthermore, the collision that generated the Occator crater (about 17.8$\pm$1.2 Ma ago, \citealt{2019Icar..318...56S}, more recently dated by \cite{2019Icar..320...60N}, who  obtained an age of 21.9$\pm$0.7 Ma) could not be the progenitor of a collisional family, as the diameter of this crater is 92 km and according to \cite{Carruba} the expected size of the progenitor is over 200 km. Thus, the presence of this band is not a reason to infer an origin of asteroid 6671 from Ceres.\\

\subsection{Asteroid 61674: The overdensity region in spectral slope density plot}\label{sec:Res:sub:Privileged_direction}

We noticed that asteroid 61674 lays in a particular place of the density plot shown in Fig. \ref{f:Res:2d_hist_shifted}, near the asymmetric dense region observed in the bottom left side of the plot. This region is mainly correlated  with cratered surface of Ceres.

\begin{figure}[!ht]
\centering
\includegraphics[width=0.49\textwidth]{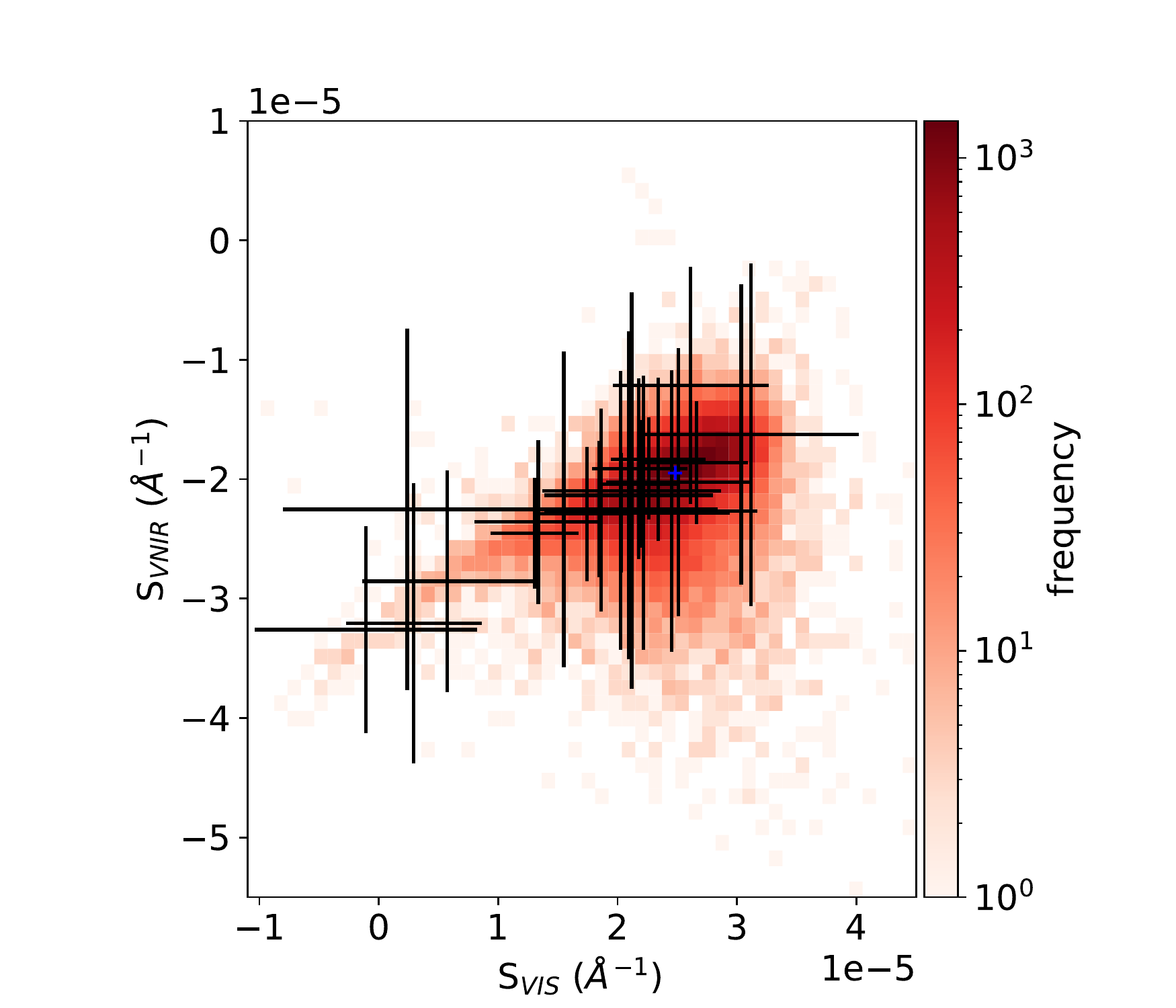}
\caption{2D density distribution of spectral slopes $S_{\rm VIS}$ and $S_{\rm VNIR}$ on the surface of Ceres, computed from the Dawn/VIR data, provided by \cite{Rousseau} and binned in $1^{\circ} \times1^{\circ} $ boxes. The median slope value for each crater from \cite{2019Icar..318...56S} and its standard deviation is represented with a black cross.}
\label{f:Res:2D_hist_crat}
\end{figure}

\begin{figure*}[!ht]
\centering
\includegraphics[width=\textwidth]{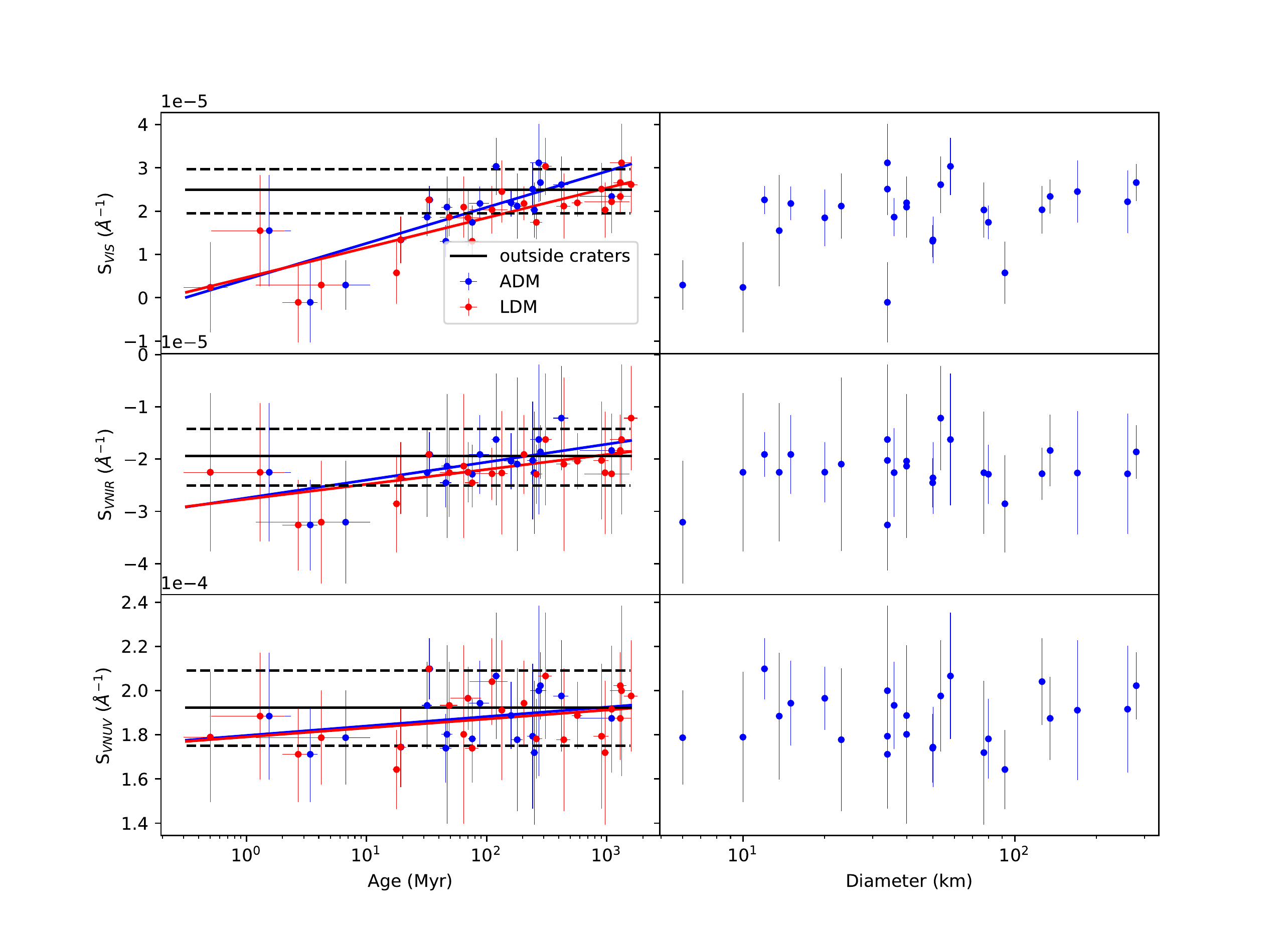}
\caption{Relation between the median spectral slopes in the three different wavelength ranges described in Section \ref{sec:dat} for craters on the surface of Ceres with their age (left) and size (right). The upper panel corresponds to $S_{\rm VIS}$, the central panel to $S_{\rm VNIR}$, and the lower panel to $S_{\rm VNUV}$. The represented ages are from two models: the asteroid derived model (in blue) and the lunar derived model (in red). The median slope value of the region outside the craters is represented with a horizontal black line. The dashed lines mark the 16th and 85th percentiles. Solid lines correspond to the least squares fit for each model.}
\label{f:Res:crater}
\end{figure*}

To study the nature of this behavior, we measured the spectral slopes of 26  different craters with measured diameters and ages based on the crater chronology collected by \cite{2019Icar..318...56S}; more information about them  is given  in Table \ref{t:Res:craters}. In order to compare these slopes with those measured in regions outside the craters, we used a larger sample of craters from  \cite{2016Sci...353.4759H}: we first selected the region inside each crater on our cylindrical projection map; we then modeled the craters as an ellipse with the semimajor axis ($a$) as the given radius in degrees and semiminor axis $b=a\times\cos(\phi)$, with $\phi$ being the latitude; finally, once all of the cratered regions were identified, we used this information to mask the map and to extract the data from noncratered surface. In Fig. \ref{f:Res:2D_hist_crat} we show the median slope for each dated crater \citep{2019Icar..318...56S} over the density plot, clearly following the overdensity region. In Fig. \ref{f:Res:crater}, we present the median value for each dated crater versus age and diameter in the three wavelength ranges defined by \cite{Rousseau} (see Section \ref{sec:dat}). The ages were obtained using two different models: the asteroid derived model (ADM) \citep{2012Sci...336..690M, 2016NatCo...712257M} and the lunar derived model (LDM) \citep{2016Sci...353.4759H}, as explained in \cite{2019Icar..318...56S}. Both models are based on crater size-frequency distribution measurements: the LDM uses direct crater observations at the surface of the Moon, and the ADM uses observations of objects in the main asteroid belt \citep{2016Sci...353.4759H}. However, both have limitations: the LDM assumes that the flux of the impacting projectiles in the asteroid belt had the same variations in the size frequency and flux of impacts as there were in the Moon, which is inconsistent with current observations; on the other hand, ADM uses only observable asteroids, bigger than $\sim$3 km radius which generate $\sim$20 km diameter craters. For smaller craters an extrapolation of the model is needed \citep{2016Sci...353.4759H}. We  include in Fig. \ref{f:Res:crater} the median slope values for the surface outside every crater in \cite{2016Sci...353.4759H}.\\

\begin{table*}[t]
\centering
\begin{tabular}{llllll} 
\hline
 Name & D (km) & Latitude ($\deg$) & Longitude ($\deg$) & LDM (Ma)& ADM (Ma)\\
 \hline
 \hline
Achita & 40 & 25.82 & 65.96 & 570 $\pm$ 60 & 160 $\pm$ 20\\
Azacca & 49.9 & -6.66 & 218.4 & 75.9 $\pm$ 10 & 45.8 $\pm$ 5 \\
Cacaguat & 13.6 & -1.2&  143.6 & 1.3 $\pm$ 0.79 & 1.55 $\pm$ 0.8 \\
Centeotl & 6.0 & 18.9 & 141.2 & 4.2 $\pm$ 3 & 6.7 $\pm$ 4 \\
Coniraya & 135 & 39.9 & 65.7 & 1300 $\pm$ 300 & 1100 $\pm$ 500 \\
Dantu & 126 & 24.3 & 138.2 & 111 $\pm$39 & - \\
Ernutet & 53.4 & 52.9&  45.5 & 1600 $\pm$ 200 & 420 $\pm$ 60 \\
Gaue & 80 & 30.8 & 86.2 & 260 $\pm$ 30 & 76 $\pm$ 6 \\
Haulani & 34 & 5.8 & 10.8 & 2.7 $\pm$ 0.7 & 3.4$\pm$ 0.5 \\
Ikapati & 50 & 33.8 & 45.6 & 19.2 $\pm$ 2.2 & 19.4 $\pm$ 1.9 \\
Kerwan & 280 & -10.8 & 124 & 1300 $\pm$160 & 281 $\pm$ 17 \\
Liber & 23 & 42.6&  37.8 & 440 $\pm$ 60 & 180 $\pm$ 20 \\
Messor & 40 & 49.9&  233.7 & 64.5 $\pm$ 2.6 & 46.8 $\pm$ 4.9 \\
Occator & 92 & 19.8 & 239.3 & 17.8 $\pm$1.2 & - \\
Omonga & 77 & 58.0 & 71.7 & 970 $\pm$ 70 & 250 $\pm$ 20 \\
Oxo & 10 & 42.2&  359.6 & 0.5 $\pm$ 0.2 & 0.5 $\pm$ 0.2 \\
Rao & 12 & 8.1 & 119.0 & 33.1 $\pm$ 2.5 & 33.6 $\pm$ 2.5 \\
Sintana & 58 & -48.1&  46.2 & 310 $\pm$ 40 & 120 $\pm$ 10 \\
Tupo & 36 & -32.3 & 88.4 & 49 $\pm$ 8 & 32 $\pm$ 3 \\
Urvara & 170 & -46.6&  249.2 & 134 $\pm$ 8 & - \\
Yalode & 260 & -42.6 & 292.5 & 1100 $\pm$ 450 & - \\
Unnamed & 34 & 39 & 247 & 906 $\pm$ 130 & 242 $\pm$ 24 \\
Unnamed & 15 & 23&  186 & 205$\pm$ 12 & 88.1 $\pm$17 \\
Unnamed & 34 & -43.3&  120.9 & 1330 $\pm$270 & 272$\pm$ 41\\
AhunaMons & 20 & -10.3 & 316.5 & 70 $\pm$ 20 & 70 $\pm$ 20 \\
\hline\\[-2mm]
\end{tabular}
\caption{Summary table with name, diameter, location, and estimated age for every crater from \cite{2019Icar..318...56S} used in this study.}
\label{t:Res:craters}
\end{table*}

In the visible range, we found a Pearson correlation coefficient between the spectral slope and $\ln(t)$ of 0.8 for both models. More specifically, the younger craters are spectrally bluer, which is consistent with the results found by \cite{2019Icar..318...56S}, who used a ratio between two filters, 0.437 $\mu$m and 0.749 $\mu$m, of the Dawn Framing Camera. Assuming that the region outside the craters is the oldest, the youngest craters have a bluer slope beyond 1$\sigma$ boundaries. This suggests a relation between the age and color on Ceres, possibly related to the space weathering effect. It is an optically nonlinear effect dependent on the exposure time according to the results obtained from laboratory experiments \citep{2006Icar..180..546B,2017Icar..285...43L}. We found a similar relation in Ceres' craters. This behavior fits the following function for each dating model:

$$S_{\rm VIS}(t_{\rm ADM})=\left(3.6\pm0.6\times\ln(t_{\rm ADM}) + 4\pm3\right)\times10^{-6} \;\;|\;\; R^2 = 0.7 \;,$$ 
$$S_{\rm VIS}(t_{\rm LDM})=\left(3.0\pm0.4\times\ln(t_{\rm LDM}) + 5\pm2\right)\times10^{-6} \;\;|\;\; R^2 = 0.7 \;.$$ 

\cite{2019Icar..318...56S} also note that invoking space weathering is mandatory for explaining this behavior. However, the composition, grain size \citep{2021Icar..35714141S}, porosity  \citep{2016Icar..267..154P, 2021NatCom..Schroeder}, and mixing modalities \citep{2018Icar..306..306R} also affect spectral slopes. Caution must therefore be exercized when interpreting these results. In the VNIR and VNUV ranges, although the fit also follows a reddening trend with age, every slope value of the craters is compatible with the 1$\sigma$ limit defined by the region outside the craters, thus, no further conclusions could therefore be reached.

It is known that large impacts have a low probability of forming \citep{2012Sci...336..690M, 2016NatCo...712257M}, so it is expected that the larger the crater, the older it is. If the spectral slope is a proxy of the crater age, the probability of there being large craters as blue as younger craters should be low. 
Following this line of reasoning, we see in the right panel of Fig. \ref{f:Res:crater} how craters with a diameter greater than 100 km are found mainly in the red part of the distribution with a median visible slope of $(2.5 \pm 0.6)\times10^{-5}$ $\AA^{-1}$. This range of slopes is in good agreement with that of the surface outside craters. Meanwhile, craters smaller than 100 km are quite spread over the color space with a bluer median visible slope of $(1.6\pm 1.0)\times10^{-5}$ $\AA^{-1}$.\\

As we have mentioned in Section \ref{sec:int}, our target asteroids are in the pristine region, have mainly been preserved since their formation, and would probably have originated through impacts that produced craters with diameters of several hundreds of kilometers \citep{Carruba}. These lines of evidence imply that the asteroids belonging to the Ceres collisional family should belong to the redder part of this overpopulation, where the largest and oldest craters are located, assuming the craters and asteroid family optically evolve in the same manner according to age. Thus, the blueness of 61674 suggests that it is less likely to be a member of a collisional family of Ceres. On the other hand, 6671 has redder slopes compatible with more weathered material. However, the presence of an absorption band, only reported in the surroundings of the young Occator crater ($17.8\pm1.2$ Ma) on the surface of Ceres, suggests that it is as fresh as this crater. Nevertheless, members of asteroid families could collide and generate a second, fresher generation of asteroids \citep{2006AJ....131.1138M}. The collisional timescale depends on the diameter of the family members; 6671 and 61674 have a diameter of 13.75 and 8.21 km, respectively \citep{Carruba}. According to \cite{2005Icar..179...63B}, the average collisional lifetime of 10 km diameter asteroids is 4.7 Ga. Following the new models for collisional disruptions in \cite{2020AJ....160...14B}, we obtained a collisional lifetime of $\sim$6 Ga. The mean age of main belt asteroids ($\mbox{T}_{\rm MB}$) depends on the collisional lifetime ($\tau_{\rm coll}$) and the late heavy bombardment age ($t_{\rm LHB}$) following the formula given by \cite{2006AJ....131.1138M}: 

$$\mbox{T}_{MB} \approx \tau_{\rm coll}\left[1-\left(1+\frac{t_{\rm LHB}}{\tau_{\rm coll}}\right)e^{-t_{\rm LHB}/\tau_{\rm coll}}\right]+t_{\rm LHB}e^{-t_{\rm LHB}/\tau_{\rm coll}}\;.$$ 

Using $\tau_{\rm coll}\in (4.7, 6)$ Ga and $t_{\rm LHB} \in (4, 4.5)$ Ga, we obtain $\mbox{T}_{\rm MB} \in (2.7,3.2)$ Ga, which is older than the highest age estimate for craters on the surface of Ceres (1.6 Ga) but of the same order of magnitude. 
Even though this line of argument indicates that those two asteroid are less likely to be part of the collisional family, one has to consider that collisional models also suggest that larger impacts may have happened in the past \citep{2016NatCo...712257M}, having been hidden by a resurfacing process.

\section{Conclusions}\label{sec:con}
Here, we have presented a spectroscopic study at visible wavelengths of 14 asteroids selected from the list of potential members of a collisional family of Ceres by \cite{Carruba}. We have reached the following conclusions:
\begin{itemize}
\item After taxonomic classification of the asteroids, comparison of their spectral slopes with that of Ceres itself, and with the distribution of slopes across the latter's surface, 12 out of the 14 asteroids are found not to be compatible with Ceres' spectra.

\item In one of the other two asteroids, 6671, we have detected an absorption band at 0.7 $\mu$m, which indicates the presence of hydrated silicates. Although there is strong evidence of hydration in Ceres, this 0.7 $\mu$m band has not been detected in any of the Dawn/VIR spectra or in the ground-based spectra of Ceres. The only detection so far is the one by \cite{2019Icar..328...69R} in the surroundings of the young Occator crater using photometric data from the Dawn Framing Camera. The band depth measured for 6671 is in agreement with that from \cite{2019Icar..328...69R}, but it is also in good agreement with the values typically observed in other primitive asteroids throughout the main belt \citep{2016A&A...586A.129M, 2018A&A...610A..25M}.

\item Our results strongly suggest that material on the surface of Ceres gets redder with time in the visible wavelength range. The blueness of the other compatible asteroid, 61674, makes it compatible with fresher crater material. 

\item We expect that asteroids with diameters of tens of kilometers that belong to a Ceres collisional family located in the pristine region are probably as red as the oldest craters. The compatibility of both 6671 and 61674 with fresh Ceres material suggest that they are not likely to be members of such a collisional family. As we should also consider the possibility that they are refreshed second generation objects, we cannot refute that this family exists or has existed. More potential members need to be observed in order to confirm or reject the existence of such a family.\\
\end{itemize}

%



\begin{acknowledgements}
FTR, JdL, ET, and JLR acknowledge financial support from the project PID2020-120464GB-I100 of the Spanish Ministerio de Ciencia e Innovación (MICINN).\\

Based on observations made with the Gran Telescopio Canarias (GTC), under observational program GTC68-17A. We especially thank David Morate for sharing his time.
\end{acknowledgements}

%
%

\bibliographystyle{aa}
\bibliography{main}

\begin{thebibliography}{54}
\expandafter\ifx\csname natexlab\endcsname\relax\def\natexlab#1{#1}\fi

\bibitem[{{Bottke} {et~al.}(2005){Bottke}, {Durda}, {Nesvorn{\'y}}, {Jedicke},
  {Morbidelli}, {Vokrouhlick{\'y}}, \& {Levison}}]{2005Icar..179...63B}
{Bottke}, W.~F., {Durda}, D.~D., {Nesvorn{\'y}}, D., {et~al.} 2005, \icarus,
  179, 63

\bibitem[{{Bottke} {et~al.}(2020){Bottke}, {Vokrouhlick{\'y}}, {Ballouz},
  {Barnouin}, {Connolly}, {Elder}, {Marchi}, {McCoy}, {Michel}, {Nolan},
  {Rizk}, {Scheeres}, {Schwartz}, {Walsh}, \& {Lauretta}}]{2020AJ....160...14B}
{Bottke}, W.~F., {Vokrouhlick{\'y}}, D., {Ballouz}, R.~L., {et~al.} 2020, \aj,
  160, 14

\bibitem[{{Brunetto} {et~al.}(2006){Brunetto}, {Romano}, {Blanco}, {Fonti},
  {Martino}, {Orofino}, \& {Verrienti}}]{2006Icar..180..546B}
{Brunetto}, R., {Romano}, F., {Blanco}, A., {et~al.} 2006, \icarus, 180, 546

\bibitem[{{Bus} \& {Binzel}(2002{\natexlab{a}})}]{2002Icar..158..146B}
{Bus}, S.~J. \& {Binzel}, R.~P. 2002{\natexlab{a}}, Icarus, 158, 146

\bibitem[{{Bus} \& {Binzel}(2002{\natexlab{b}})}]{2002Icar..158..106B}
{Bus}, S.~J. \& {Binzel}, R.~P. 2002{\natexlab{b}}, Icarus, 158, 106

\bibitem[{{Carrozzo} {et~al.}(2016){Carrozzo}, {Raponi}, {De Sanctis},
  {Ammannito}, {Giardino}, {D'Aversa}, {Fonte}, \&
  {Tosi}}]{2016RScI...87l4501C}
{Carrozzo}, F.~G., {Raponi}, A., {De Sanctis}, M.~C., {et~al.} 2016, Review of
  Scientific Instruments, 87, 124501

\bibitem[{{Carruba} {et~al.}(2016){Carruba}, {Nesvorn{\'y}}, {Marchi}, \&
  {Aljbaae}}]{Carruba}
{Carruba}, V., {Nesvorn{\'y}}, D., {Marchi}, S., \& {Aljbaae}, S. 2016, Monthly
  Notices of the Royal Astronomical Society, 458, 1117

\bibitem[{{Cellino} {et~al.}(2002){Cellino}, {Bus}, {Doressoundiram}, \&
  {Lazzaro}}]{Cellino2002}
{Cellino}, A., {Bus}, S.~J., {Doressoundiram}, A., \& {Lazzaro}, D. 2002,
  {Spectroscopic Properties of Asteroid Families} (University of Arizona Press,
  Tucson), 633--643

\bibitem[{{Cepa}(2010)}]{2010ASSP...14...15C}
{Cepa}, J. 2010, Astrophysics and Space Science Proceedings, 14, 15

\bibitem[{{Cepa} {et~al.}(2000){Cepa}, {Aguiar}, {Escalera},
  {Gonzalez-Serrano}, {Joven-Alvarez}, {Peraza}, {Rasilla}, {Rodriguez-Ramos},
  {Gonzalez}, {Cobos Duenas}, {Sanchez}, {Tejada}, {Bland-Hawthorn},
  {Militello}, \& {Rosa}}]{2000SPIE.4008..623C}
{Cepa}, J., {Aguiar}, M., {Escalera}, V.~G., {et~al.} 2000, in Society of
  Photo-Optical Instrumentation Engineers (SPIE) Conference Series, Vol. 4008,
  Optical and IR Telescope Instrumentation and Detectors, ed. M.~{Iye} \& A.~F.
  {Moorwood}, 623--631

\bibitem[{{Chapman} \& {Gaffey}(1979)}]{1979aste.book.1064C}
{Chapman}, C.~R. \& {Gaffey}, M.~J. 1979, {Spectral reflectances of the
  asteroids}, ed. T.~{Gehrels} \& M.~S. {Matthews}, 1064--1089

\bibitem[{{Ciarniello} {et~al.}(2015){Ciarniello}, {Capaccioni}, {Filacchione},
  {Raponi}, {Tosi}, {De Sanctis}, {Capria}, {Erard}, {Bockelee-Morvan},
  {Leyrat}, {Arnold}, {Barucci}, {Beck}, {Bellucci}, {Fornasier}, {Longobardo},
  {Mottola}, {Palomba}, {Quirico}, \& {Schmitt}}]{2015A&A...583A..31C}
{Ciarniello}, M., {Capaccioni}, F., {Filacchione}, G., {et~al.} 2015, \aap,
  583, A31

\bibitem[{{Ciarniello} {et~al.}(2017){Ciarniello}, {De Sanctis}, {Ammannito},
  {Raponi}, {Longobardo}, {Palomba}, {Carrozzo}, {Tosi}, {Li}, {Schr{\"o}der},
  {Zambon}, {Frigeri}, {Fonte}, {Giardino}, {Pieters}, {Raymond}, \&
  {Russell}}]{2017A&A...598A.130C}
{Ciarniello}, M., {De Sanctis}, M.~C., {Ammannito}, E., {et~al.} 2017,
  Astronomy \& Astrophysics, 598, A130

\bibitem[{{de Sanctis} {et~al.}(2015){de Sanctis}, {Ammannito}, {Raponi},
  {Marchi}, {McCord}, {McSween}, {Capaccioni}, {Capria}, {Carrozzo},
  {Ciarniello}, {Longobardo}, {Tosi}, {Fonte}, {Formisano}, {Frigeri},
  {Giardino}, {Magni}, {Palomba}, {Turrini}, {Zambon}, {Combe}, {Feldman},
  {Jaumann}, {McFadden}, {Pieters}, {Prettyman}, {Toplis}, {Raymond}, \&
  {Russell}}]{2015Natur.528..241D}
{de Sanctis}, M.~C., {Ammannito}, E., {Raponi}, A., {et~al.} 2015, Nature, 528,
  241

\bibitem[{{de Sanctis} {et~al.}(2011){de Sanctis}, {Coradini}, {Ammannito},
  {Filacchione}, {Capria}, {Fonte}, {Magni}, {Barbis}, {Bini}, {Dami},
  {Ficai-Veltroni}, \& {Preti}}]{2011SSRv..163..329D}
{de Sanctis}, M.~C., {Coradini}, A., {Ammannito}, E., {et~al.} 2011, \ssr, 163,
  329

\bibitem[{{DeMeo} {et~al.}(2009){DeMeo}, {Binzel}, {Slivan}, \&
  {Bus}}]{2009Icar..202..160D}
{DeMeo}, F.~E., {Binzel}, R.~P., {Slivan}, S.~M., \& {Bus}, S.~J. 2009, Icarus,
  202, 160

\bibitem[{{Fornasier} {et~al.}(2014){Fornasier}, {Lantz}, {Barucci}, \&
  {Lazzarin}}]{2014Icar..233..163F}
{Fornasier}, S., {Lantz}, C., {Barucci}, M.~A., \& {Lazzarin}, M. 2014, Icarus,
  233, 163

\bibitem[{{Fornasier} {et~al.}(1999){Fornasier}, {Lazzarin}, {Barbieri}, \&
  {Barucci}}]{1999A&AS..135...65F}
{Fornasier}, S., {Lazzarin}, M., {Barbieri}, C., \& {Barucci}, M.~A. 1999,
  Astronomy \& Astrophysics Supplement, 135, 65

\bibitem[{{Fujiya} {et~al.}(2019){Fujiya}, {Hoppe}, {Ushikubo}, {Fukuda},
  {Lindgren}, {Lee}, {Koike}, {Shirai}, \& {Sano}}]{2019NatAs...3..910F}
{Fujiya}, W., {Hoppe}, P., {Ushikubo}, T., {et~al.} 2019, Nature Astronomy, 3,
  910

\bibitem[{{Hiesinger} {et~al.}(2016){Hiesinger}, {Marchi}, {Schmedemann},
  {Schenk}, {Pasckert}, {Neesemann}, {O'Brien}, {Kneissl}, {Ermakov}, {Fu},
  {Bland}, {Nathues}, {Platz}, {Williams}, {Jaumann}, {Castillo-Rogez},
  {Ruesch}, {Schmidt}, {Park}, {Preusker}, {Buczkowski}, {Russell}, \&
  {Raymond}}]{2016Sci...353.4759H}
{Hiesinger}, H., {Marchi}, S., {Schmedemann}, N., {et~al.} 2016, Science, 353,
  aaf4758

\bibitem[{{Knezevic} {et~al.}(2002){Knezevic}, {Lema{\^\i}tre}, \&
  {Milani}}]{Knezevic2002}
{Knezevic}, Z., {Lema{\^\i}tre}, A., \& {Milani}, A. 2002, in Asteroids III,
  ed. W.~F.~J. {Bottke}, A.~{Cellino}, P.~{Paolicchi}, \& R.~P. {Binzel}
  (University of Arizona Press, Tucson), 603--612

\bibitem[{{Lantz} {et~al.}(2017){Lantz}, {Brunetto}, {Barucci}, {Fornasier},
  {Baklouti}, {Bour{\c{c}}ois}, \& {Godard}}]{2017Icar..285...43L}
{Lantz}, C., {Brunetto}, R., {Barucci}, M.~A., {et~al.} 2017, \icarus, 285, 43

\bibitem[{{Lazzaro} {et~al.}(2004){Lazzaro}, {Angeli}, {Carvano},
  {Moth{\'e}-Diniz}, {Duffard}, \& {Florczak}}]{2004Icar..172..179L}
{Lazzaro}, D., {Angeli}, C.~A., {Carvano}, J.~M., {et~al.} 2004, Icarus, 172,
  179

\bibitem[{{Lebofsky} {et~al.}(1981){Lebofsky}, {Feierberg}, {Tokunaga},
  {Larson}, \& {Johnson}}]{1981Icar...48..453L}
{Lebofsky}, L.~A., {Feierberg}, M.~A., {Tokunaga}, A.~T., {Larson}, H.~P., \&
  {Johnson}, J.~R. 1981, \icarus, 48, 453

\bibitem[{{Marchi} {et~al.}(2016){Marchi}, {Ermakov}, {Raymond}, {Fu},
  {O'Brien}, {Bland}, {Ammannito}, {de Sanctis}, {Bowling}, {Schenk}, {Scully},
  {Buczkowski}, {Williams}, {Hiesinger}, \& {Russell}}]{2016NatCo...712257M}
{Marchi}, S., {Ermakov}, A.~I., {Raymond}, C.~A., {et~al.} 2016, Nature
  Communications, 7, 12257

\bibitem[{{Marchi} {et~al.}(2012){Marchi}, {McSween}, {O'Brien}, {Schenk}, {De
  Sanctis}, {Gaskell}, {Jaumann}, {Mottola}, {Preusker}, {Raymond}, {Roatsch},
  \& {Russell}}]{2012Sci...336..690M}
{Marchi}, S., {McSween}, H.~Y., {O'Brien}, D.~P., {et~al.} 2012, Science, 336,
  690

\bibitem[{{Marchi} {et~al.}(2006){Marchi}, {Paolicchi}, {Lazzarin}, \&
  {Magrin}}]{2006AJ....131.1138M}
{Marchi}, S., {Paolicchi}, P., {Lazzarin}, M., \& {Magrin}, S. 2006, \aj, 131,
  1138

\bibitem[{{McFadden} {et~al.}(1984){McFadden}, {Gaffey}, \&
  {McCord}}]{1984Icar...59...25M}
{McFadden}, L.~A., {Gaffey}, M.~J., \& {McCord}, T.~B. 1984, \icarus, 59, 25

\bibitem[{{Milani} {et~al.}(2014){Milani}, {Cellino}, {Kne{\v{z}}evi{\'c}},
  {Novakovi{\'c}}, {Spoto}, \& {Paolicchi}}]{2014Icar..239...46M}
{Milani}, A., {Cellino}, A., {Kne{\v{z}}evi{\'c}}, Z., {et~al.} 2014, \icarus,
  239, 46

\bibitem[{{Morate} {et~al.}(2018){Morate}, {de Le{\'o}n}, {De Pr{\'a}},
  {Licandro}, {Cabrera-Lavers}, {Campins}, \&
  {Pinilla-Alonso}}]{2018A&A...610A..25M}
{Morate}, D., {de Le{\'o}n}, J., {De Pr{\'a}}, M., {et~al.} 2018, Astronomy \&
  Astrophysics, 610, A25

\bibitem[{{Morate} {et~al.}(2016){Morate}, {de Le{\'o}n}, {De Pr{\'a}},
  {Licandro}, {Cabrera-Lavers}, {Campins}, {Pinilla-Alonso}, \&
  {Al{\'\i}-Lagoa}}]{2016A&A...586A.129M}
{Morate}, D., {de Le{\'o}n}, J., {De Pr{\'a}}, M., {et~al.} 2016, Astronomy \&
  Astrophysics, 586, A129

\bibitem[{{Nathues} {et~al.}(2016){Nathues}, {Hoffmann}, {Platz}, {Thangjam},
  {Cloutis}, {Reddy}, {Le Corre}, {Li}, {Mengel}, {Rivkin}, {Applin},
  {Schaefer}, {Christensen}, {Sierks}, {Ripken}, {Schmidt}, {Hiesinger},
  {Sykes}, {Sizemore}, {Preusker}, \& {Russell}}]{2016P&SS..134..122N}
{Nathues}, A., {Hoffmann}, M., {Platz}, T., {et~al.} 2016, \planss, 134, 122

\bibitem[{{Neesemann} {et~al.}(2019){Neesemann}, {van Gasselt}, {Schmedemann},
  {Marchi}, {Walter}, {Preusker}, {Michael}, {Kneissl}, {Hiesinger}, {Jaumann},
  {Roatsch}, {Raymond}, \& {Russell}}]{2019Icar..320...60N}
{Neesemann}, A., {van Gasselt}, S., {Schmedemann}, N., {et~al.} 2019, \icarus,
  320, 60

\bibitem[{{Nesvorn{\'y}} {et~al.}(2015){Nesvorn{\'y}}, {Bro{\v{z}}}, \&
  {Carruba}}]{2015aste.book..297N}
{Nesvorn{\'y}}, D., {Bro{\v{z}}}, M., \& {Carruba}, V. 2015, {Identification
  and Dynamical Properties of Asteroid Families}, 297--321

\bibitem[{{Perna} {et~al.}(2015){Perna}, {Ka{\v{n}}uchov{\'a}}, {Ieva},
  {Fornasier}, {Barucci}, {Lantz}, {Dotto}, \&
  {Strazzulla}}]{2015A&A...575L...1P}
{Perna}, D., {Ka{\v{n}}uchov{\'a}}, Z., {Ieva}, S., {et~al.} 2015, \aap, 575,
  L1

\bibitem[{{Poch} {et~al.}(2016){Poch}, {Pommerol}, {Jost}, {Carrasco}, {Szopa},
  \& {Thomas}}]{2016Icar..267..154P}
{Poch}, O., {Pommerol}, A., {Jost}, B., {et~al.} 2016, \icarus, 267, 154

\bibitem[{{Popescu} {et~al.}(2012){Popescu}, {Birlan}, \& {Nedelcu}}]{m4ast}
{Popescu}, M., {Birlan}, M., \& {Nedelcu}, D.~A. 2012, Astronomy \&
  Astrophysics, 544, A130

\bibitem[{{Reddy} {et~al.}(2015){Reddy}, {Li}, {Gary}, {Sanchez}, {Stephens},
  {Megna}, {Coley}, {Nathues}, {Le Corre}, \& {Hoffmann}}]{2015Icar..260..332R}
{Reddy}, V., {Li}, J.-Y., {Gary}, B.~L., {et~al.} 2015, \icarus, 260, 332

\bibitem[{{Rivkin}(2012)}]{2012Icar..221..744R}
{Rivkin}, A.~S. 2012, Icarus, 221, 744

\bibitem[{{Rivkin} {et~al.}(2006){Rivkin}, {Volquardsen}, \&
  {Clark}}]{2006Icar..185..563R}
{Rivkin}, A.~S., {Volquardsen}, E.~L., \& {Clark}, B.~E. 2006, \icarus, 185,
  563

\bibitem[{{Rizos} {et~al.}(2019){Rizos}, {de Le{\'o}n}, {Licandro}, {Campins},
  {Popescu}, {Pinilla-Alonso}, {Golish}, {de Pr{\'a}}, \&
  {Lauretta}}]{2019Icar..328...69R}
{Rizos}, J.~L., {de Le{\'o}n}, J., {Licandro}, J., {et~al.} 2019, Icarus, 328,
  69

\bibitem[{{Rousseau} {et~al.}(2020){Rousseau}, {De Sanctis}, {Raponi},
  {Ciarniello}, {Ammannito}, {Frigeri}, {Ferrari}, {De Angelis}, {Carrozzo},
  {Tosi}, {Schr{\"o}der}, {Raymond}, \& {Russell}}]{Rousseau}
{Rousseau}, B., {De Sanctis}, M.~C., {Raponi}, A., {et~al.} 2020, \aap, 642,
  A74

\bibitem[{{Rousseau} {et~al.}(2018){Rousseau}, {{\'E}rard}, {Beck}, {Quirico},
  {Schmitt}, {Brissaud}, {Montes-Hernandez}, {Capaccioni}, {Filacchione},
  {Bockel{\'e}e-Morvan}, {Leyrat}, {Ciarniello}, {Raponi}, {Kappel}, {Arnold},
  {Moroz}, {Palomba}, {Tosi}, \& {Virtis Team}}]{2018Icar..306..306R}
{Rousseau}, B., {{\'E}rard}, S., {Beck}, P., {et~al.} 2018, \icarus, 306, 306

\bibitem[{{Schr{\"o}der} {et~al.}(2021){Schr{\"o}der}, {Poch}, {Ferrari}, {De
  Angelis}, {Sultana}, {Potin}, {Beck}, {De Sanctis}, , \&
  {Schmitt}}]{2021NatCom..Schroeder}
{Schr{\"o}der}, S.~E., {Poch}, O., {Ferrari}, M., {et~al.} 2021, Nature
  Communications, 12

\bibitem[{{Stephan} {et~al.}(2019){Stephan}, {Jaumann}, {Zambon}, {Carrozzo},
  {Wagner}, {Longobardo}, {Palomba}, {De Sanctis}, {Tosi}, {Ammannito},
  {Combe}, {Mc Fadden}, {Krohn}, {Schulzeck}, {von der Gathen}, {Williams},
  {Scully}, {Schmedemann}, {Neesemann}, {Roatsch}, {Matz}, {Preusker},
  {Raymond}, \& {Russell}}]{2019Icar..318...56S}
{Stephan}, K., {Jaumann}, R., {Zambon}, F., {et~al.} 2019, \icarus, 318, 56

\bibitem[{{Sultana} {et~al.}(2021){Sultana}, {Poch}, {Beck}, {Schmitt}, \&
  {Quirico}}]{2021Icar..35714141S}
{Sultana}, R., {Poch}, O., {Beck}, P., {Schmitt}, B., \& {Quirico}, E. 2021,
  \icarus, 357, 114141

\bibitem[{{Usui} {et~al.}(2019){Usui}, {Hasegawa}, {Ootsubo}, \&
  {Onaka}}]{2019PASJ...71....1U}
{Usui}, F., {Hasegawa}, S., {Ootsubo}, T., \& {Onaka}, T. 2019, \pasj, 71, 1

\bibitem[{{Valsecchi} {et~al.}(1989){Valsecchi}, {Carusi}, {Knezevic},
  {Kresak}, \& {Williams}}]{1989aste.conf..368V}
{Valsecchi}, G.~B., {Carusi}, A., {Knezevic}, Z., {Kresak}, L., \& {Williams},
  J.~G. 1989, in Asteroids II, ed. R.~P. {Binzel}, T.~{Gehrels}, \& M.~S.
  {Matthews}, 368--385

\bibitem[{{Vernazza} {et~al.}(2005){Vernazza}, {Moth{\'e}-Diniz}, {Barucci},
  {Birlan}, {Carvano}, {Strazzulla}, {Fulchignoni}, \&
  {Migliorini}}]{2005A&A...436.1113V}
{Vernazza}, P., {Moth{\'e}-Diniz}, T., {Barucci}, M.~A., {et~al.} 2005, \aap,
  436, 1113

\bibitem[{{Vilas}(1994)}]{1994Icar..111..456V}
{Vilas}, F. 1994, Icarus, 111, 456

\bibitem[{{Vilas} \& {Gaffey}(1989)}]{1989Sci...246..790V}
{Vilas}, F. \& {Gaffey}, M.~J. 1989, Science, 246, 790

\bibitem[{{Vilas} {et~al.}(1993){Vilas}, {Larson}, {Hatch}, \&
  {Jarvis}}]{Vilas1993}
{Vilas}, F., {Larson}, S.~M., {Hatch}, E.~C., \& {Jarvis}, K.~S. 1993, \icarus,
  105, 67

\bibitem[{{Vilas} \& {McFadden}(1992)}]{1992Icar..100...85V}
{Vilas}, F. \& {McFadden}, L.~A. 1992, \icarus, 100, 85

\bibitem[{{Zellner} {et~al.}(1985){Zellner}, {Tholen}, \&
  {Tedesco}}]{1985Icar...61..355Z}
{Zellner}, B., {Tholen}, D.~J., \& {Tedesco}, E.~F. 1985, Icarus, 61, 355

\end{thebibliography}

\end{document}